\documentclass[runningheads]{llncs}
\usepackage{graphicx}
\usepackage[utf8]{inputenc}
\usepackage{todonotes}
\usepackage{todonotes}
\usepackage{soul}
\usepackage{url}
\usepackage{hyperref}
\usepackage{paralist}
\usepackage{tabularx}

\begin{document}
\title{Living Lab Evaluation for Life and Social Sciences Search Platforms - LiLAS at CLEF 2021}

\titlerunning{Living Lab Evaluation for Life and Social Sciences Search Platforms}
%

\author{Philipp Schaer\inst{1}\orcidID{0000-0002-8817-4632} 
\and
Johann Schaible\inst{2}\orcidID{0000-0002-5441-7640} 
\and
Leyla Jael Castro \inst{3}\orcidID{0000-0003-3986-0510}
}
\authorrunning{Schaer, Schaible, and Garcia Castro}
%
\institute{
TH Köln - University of Applied Sciences, Germany\\ 
\email{philipp.schaer@th-koeln.de}\\
\and
GESIS - Leibniz Institute for the Social Sciences, Germany
\email{johann.schaible@gesis.org}\\ 
\and
ZB MED - Information Centre for Life Sciences, Germany\\
\email{ljgarcia@zbmed.de}
}
\maketitle              
\begin{abstract}
Meta-evaluation studies of system performances in controlled offline evaluation campaigns, like TREC and CLEF, show a need for innovation in evaluating IR-systems. The field of academic search is no exception to this.
This might be related to the fact that relevance in academic search is multilayered and therefore the aspect of user-centric evaluation is becoming more and more important.
The Living Labs for Academic Search (LiLAS) lab aims to strengthen the concept of user-centric living labs for the domain of academic search by allowing participants to evaluate their retrieval approaches in two real-world academic search systems from the life sciences and the social sciences.
To this end, we provide participants with metadata on the systems' content as well as candidate lists with the task to rank the most relevant candidate to the top. Using the STELLA-infrastructure, we allow participants to easily integrate their approaches into the real-world systems and provide the possibility to compare different approaches at the same time.


\end{abstract}
%
%
%

\section{Introduction and Background}

Scientific information and knowledge are growing at an exponential rate \cite{de_solla_price_little_1963}. This includes not only the traditional journal publication but also a vast amount of preprints, research data sets, code, survey data, and many others research objects. This heterogeneity and mass of documents and data sets introduces new challenges to the disciplines of information retrieval (IR), recommender systems, digital libraries, or more generally the field of academic search systems. Progress in these fields is usually evaluated by means of shared tasks that are based on the principles of Cranfield/TREC-style studies. Typical shared tasks at CLEF and TREC are based on the offline computation of results/runs missing a valuable link to real-world environments~\cite{ferro_continuous_2019}. Most recently the TREC-COVID \cite{DBLP:journals/corr/abs-2005-04474} evaluation campaign run by NIST attracted a high number of participants and showed the high impact of scientific retrieval tasks in the community. 

TREC-COVID showed the massive retrieval performance that recent deep learning approaches are capable of; however, classic vector-space retrieval using the SMART system was also highly successful \footnote{\url{https://ir.nist.gov/covidSubmit/archive.html}}. This can be attributed to the limitations of the test collection based evaluation approach of TREC-COVID and the general need for innovation in the field of academic search and IR.  Meta-evaluation studies of system performances in controlled offline evaluation campaigns, like TREC and CLEF, show a need for innovation in evaluating IR-systems~\cite{yang_critically_2019,armstrong_improvements_2009}. The field of academic search is no exception to this. The central concern of academic search is to find both relevant and high-quality documents. The question of what constitutes relevance in academic search is multilayered \cite{DBLP:conf/ecir/CarevicS14} and an ongoing research area.

To compensate for these shortcomings, e.g., lack of extensive training corpora, the living labs concept was introduced. It is meant to carry out online evaluations within a fixed methodological, organizational, and technical framework and open them up to other actors. As in a laboratory TREC environment, the aim is to prevent as many disturbing influences as possible while at the same time preserving the advantages of evaluation in live systems. For example, in the context of online IR experiments, document inventories that change over time can cause problems. Furthermore, it must be ensured that all experimental rankings have the chance to be calculated and displayed regardless of the system performance. These and other factors can be taken into account in living labs and thus form a bridge between structured, but also rigid, laboratory evaluation and free, but also less planned and controlled online evaluation. The best-known case of a living lab became known in 2014 as the ``Facebook Experiment'' \cite{kramer_experimental_2014}. Even though this experiment was pervasive and did not serve the system's direct evaluation, it shows the possibilities that living labs and online experiments offer. Unfortunately, these experiments are limited by the actual access to the systems itself. As long as one is not the operator of a large-scale platform, one is most likely unable to perform such experiments. Therefore previous attempts such as Living Labs for Information Retrieval (LL4IR) and Open Search initiatives~\cite{balog2016overview} were established at CLEF or TREC to bring together IR researchers and platform operators. Likewise, the NewsREEL workshop series was an attempt to do the same for recommendation systems~\cite{lommatzsch2018newsreel}.
In summary, the living lab evaluation paradigm represents a user-centered study methodology for researchers to evaluate the performance of retrieval systems within real-world applications. As such, it offers a more realistic experiment and evaluation environment than offline test collections, and therefore should be further investigated to raise IR-evaluation to the next level. 

Fuhr \cite{Fuhr:2018:CMI:3190580.3190586} argues that evaluation initiatives should take a leading role in improving the current IR evaluation practice. We want to move beyond the traditional offline evaluation setup and bring together industry and practice evaluation techniques into the academic realm. Therefore, utilizing online evaluations, taking the actual user into account, would be a step forward towards improving the evaluation landscape. Following these lines, the primary motivation behind the LiLAS (Living Labs for Academic Search) lab at CLEF 2021 is to learn more about 

\begin{itemize}
    \item the potentials and limitations of different styles of living labs for search evaluation. Here we would like to compare pre-computed results and those provided by our live evaluation framework STELLA \cite{breuer2019} which incorporates interactions with end-users.
    \item the reproducibility of click-based evaluations in the academic domain. 
    \item the validity and expressiveness of click-based and relevance assessment-based evaluation metrics on small to medium-scale academic platforms.
\end{itemize}

After LiLAS ran as a workshop lab at CLEF 2020 \cite{DBLP:conf/clef/SchaerSC20,DBLP:conf/ecir/SchaerSM20}, in 2021 a full evaluation lab will take place. This lab's unique selling point is that we offer two tasks to test this approach in two different academic search domains and evaluation setups.

\section{Evaluation Infrastructure and Submission Types}
\label{sec:infrastructure}

\begin{figure}[t]
    \centering
    \includegraphics[width=\linewidth]{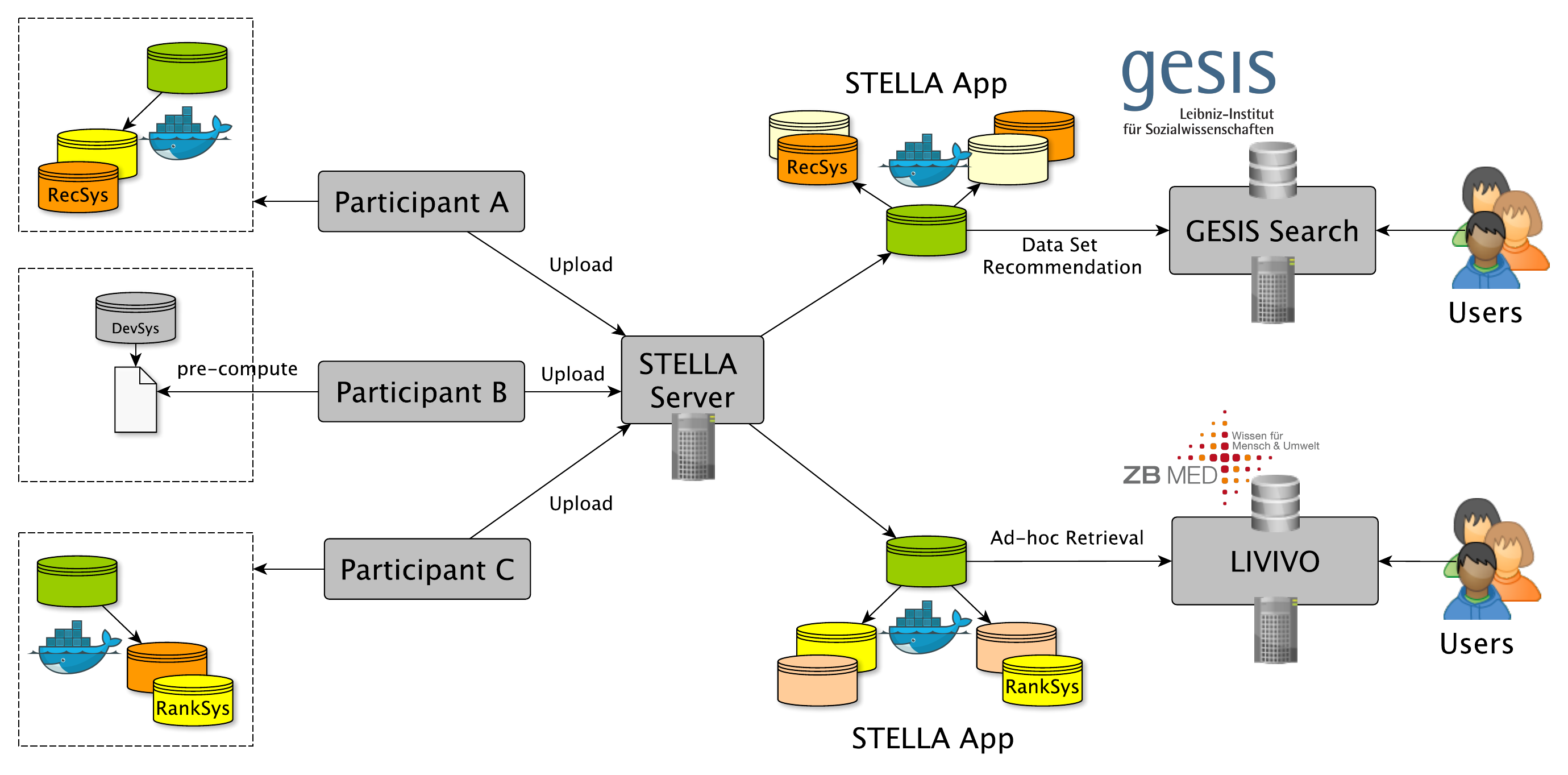}
    \caption{Schematic representation of the two core tasks: Adhoc search and dataset recommendation in the two academic search systems GESIS Search and LIVIVO. Participants can upload pre-computed results or share their systems in form of Docker containers for both tasks to the STELLA Server, which distributes the experimental data and systems to the STELLA App on premise of GESIS Search and LIVIVO. This app curates the evaluation process by offering A/B-testing or interleaving setups. The results of the runs are aggregated and displayed in a dashboard on the STELLA Server.}
    \label{fig:infrastructure}
\end{figure}

Based on previous work done in campaigns such as LL4IR at CLEF and Open Search at TREC, we built a living lab evaluation infrastructure named STELLA currently integrated into two academic search systems to allow a more realistic evaluation setup \cite{breuer2019}. 
In addition to the previous labs, where pre-computed results were submitted, we now offer participants the possibility to submit Docker containers that can be run within STELLA. This simple but yet powerful mechanism enhances the evaluation state of the art. 
Figure~\ref{fig:infrastructure} shows the schematics of STELLA in the context of this evaluation lab, which we describe in the following.

LiLAS offers two different evaluation tasks: \emph{Academic ad-hoc retrieval} for the multi-lingual and multi-source Life Science search portal LIVIVO and \emph{research data recommendation} within the Social Science portal GESIS Search.
For both tasks, participants are invited to submit
\begin{itemize}
\item \textbf{Type A} \emph{pre-computed runs} based on previously compiled queries (ad-hoc search) or documents (research data recommendations) from server logs (comparable to the CLEF LL4IR or TREC Open Search labs~\cite{DBLP:journals/jdiq/JagermanBR18}) or 
\item \textbf{Type B} \emph{Docker containers} of full running retrieval/recommendation systems that run within our evaluation framework called STELLA. 
\end{itemize}
For type A, participants pre-compute result files following TREC run file syntax and submit them for integration into the live systems. For type B, participants encapsulate their retrieval system into a Docker container following some simple implementation rules inspired by the OSIRRC workshop at SIGIR 2019. 
We release datasets containing queries and metadata of documents and data sets from the two systems mentioned above for training purposes. We offer a list of candidate documents and candidate research data for each query and seed document, respectively, so participants focus on the actual ranking approaches behind the ad-hoc search and recommendation task. 




\section{Task 1: Ad-hoc Search Ranking}
\noindent\textbf{Motivation.} Finding the most relevant publications to a query remains a challenge in scholarly Information Retrieval systems, even more in multi-lingual and cross-domain environments. 

\noindent\textbf{Task description.} The participants are asked to define and implement their ranking approach for a multi-lingual candidate documents list. A good ranking should present users with the most relevant documents regarding a query on top of the result set. Regardless of the language used to pose the query, the retrieval can include candidate documents in multiple languages. Participants can submit type A and type B results. Type A rankings will be based on the most common queries in LIVIVO (see below), while Type B submissions should rank candidate documents for any incoming query. 

\noindent\textbf{Dataset and lab data.} LIVIVO\footnote{\url{https://livivo.de}} is a search portal developed by ZB MED~- Information Centre for Life Sciences\footnote{\url{https://zbmed.de}}, providing comprehensive access to literature in life sciences, including resources from medicine, health, environment, agriculture, and nutrition. LIVIVO corpus consists of about 80 million documents from more than 50 data sources in multiple languages (e.g., English, German, French) covering various scholarly publication types (e.g., conferences, preprints, peer-review journals). We provide LiLAS participants with a set of common queries with their corresponding candidate documents (and metadata), in which participants will rank according to their relevance regarding the query. We also provide participants with logs regarding which users have accessed documents and in what order (e.g., click-through rate, if available). 

\noindent\textbf{Evaluation}
Participating approaches will be evaluated in the LIVIVO production system on gains, losses, and ties regarding user preferences (e.g., click-through rate per query and candidates list). We will follow a Team Draft Interleaving (TDI) approach \cite{radlinski_how_2008,schuth_extended_2015} where LIVIVO and participants rankings are interleaved and presented together. This way, we will be able to compare all participating systems against each other.


\section{Task 2: Research Data Recommendations}
\noindent\textbf{Motivation.} 
Research data is of high importance in scientific research, especially when making progress in experimental investigations. 
However, finding useful research data can be difficult and cumbersome, even if using dataset search engines, such as Google Dataset Search\footnote{\url{https://datasetsearch.research.google.com/}}.
Therefore, one possible solution is to recommend ``appropriate'' research data sets to users based on research articles, i.e., publications, of the users' interest.  

\noindent\textbf{Task description.}
The main task here is to provide recommendations of research data that are \emph{relevant} to the publication the user is currently viewing. 
For example, the user is interested in the impact of religion on political elections. She finds a publication regarding that topic, which has a set of research data candidates covering the same topic. The task is to rank the most relevant candidates to the top.
Participants can submit type A and type B results. 
Whereas the pre-computed type A results comprise recommendations only for publications and research data existing in the provided lab data, the Docker variant in type B will also compute recommendations for publications and research data that have recently been added to the real-life search system.

\noindent\textbf{Dataset and lab data.}
The data for this task is taken from the academic search system \emph{GESIS Search}\footnote{\url{https://search.gesis.org/}}~\cite{DBLP:conf/jcdl/HienertKBZM19}. Besides social science literature (107k publications), it also provides research data (77k) on social science topics,
out of which the participants are given the metadata to all publications and all contained research data.  
The publications are mostly in English and German and are annotated with further textual metadata like title, abstract, topic, persons, and others. 
Metadata on research data comprises (among others) a title, topics, datatype, abstract, collection method and universe, temporal and geographical coverage, primary investigators, as well as contributors in English and/or German.
The set of research data candidates for each publication, which is given to the participants as well, is computed based on context similarity between publications and research data. 

\noindent\textbf{Evaluation.}
With an A/B-testing, the GESIS Search users will be shown the recommendations separated by the users' session-id. This means, for each session-id, STELLA selects one recommendation approach out of all participants. This way, we are able to compare all participating systems against each other without confusing the user with different recommendations for the same publication. 
In both type A and type B, the participating approaches will be evaluated in the GESIS Search productive system, where the top-$k$ (with $3 \leq k \leq 10$) recommendations are shown to the user.
The evaluation itself is performed using implicit and explicit feedback. 
For the implicit feedback, we calculate the click-through-rate (CTR) as well as the bounce rate once a user clicks on a recommended dataset. 
The explicit feedback is gathered via options, such as a \emph{thumbs up} and \emph{thumbs down}, in which the users can indicate whether the recommendation was relevant to them or not. 
\section{Conclusion}

Academic Search is a timeless research domain and a very recent topic, as shown by TREC-COVID. We want to support the latest research in this field by bringing together real-world platforms and academic researchers on a joint living lab research infrastructure. As the search for scientific material is more than just ``10 blue links'', we see the demand for domain-specific retrieval tasks, which comprises document, dataset as well as bibliometric-enhanced retrieval. To allow academic researchers to go beyond proprietary web-search platforms like Google Scholar (or Google Dataset Search) or digital libraries like the ACM Digital Library, we focus on mid-size scientific search systems. The two tasks we run in 2021 are a starting point and are designed to be as open as possible by offering the possibility to submit both pre-computed and Docker-based results and systems. 
In the future, STELLA can be exploited and integrated into other academic search systems enabling online evaluations in different domains.

\bibliographystyle{splncs04}
\bibliography{ecir2021-lilas}

\begin{thebibliography}{10}
\providecommand{\url}[1]{\texttt{#1}}
\providecommand{\urlprefix}{URL }
\providecommand{\doi}[1]{https://doi.org/#1}

\bibitem{armstrong_improvements_2009}
Armstrong, T.G., Moffat, A., Webber, W., Zobel, J.: Improvements that don't add
  up: ad-hoc retrieval results since 1998. In: Proceeding of the 18th {ACM}
  conference on information and knowledge management. pp. 601--610. {CIKM} '09,
  ACM, Hong Kong, China (2009). \doi{10.1145/1645953.1646031}

\bibitem{balog2016overview}
Balog, K., Schuth, A., Dekker, P., Schaer, P., Tavakolpoursaleh, N., Chuang,
  P.Y.: Overview of the trec 2016 open search track. In: Proceedings of the
  Twenty-Fifth Text REtrieval Conference (TREC 2016). NIST (2016)

\bibitem{breuer2019}
Breuer, T., Schaer, P., Tavakolpoursaleh, N., Schaible, J., Wolff, B.,
  M{\"{u}}ller, B.: {STELLA:} towards a framework for the reproducibility of
  online search experiments. In: Clancy, R., Ferro, N., Hauff, C., Lin, J.,
  Sakai, T., Wu, Z.Z. (eds.) Proceedings of the Open-Source {IR} Replicability
  Challenge co-located with 42nd International {ACM} {SIGIR} Conference on
  Research and Development in Information Retrieval, OSIRRC@SIGIR 2019, Paris,
  France, July 25, 2019. {CEUR} Workshop Proceedings, vol.~2409, pp. 8--11.
  CEUR-WS.org (2019), \url{http://ceur-ws.org/Vol-2409/position01.pdf}

\bibitem{DBLP:conf/ecir/CarevicS14}
Carevic, Z., Schaer, P.: On the connection between citation-based and topical
  relevance ranking: Results of a pretest using isearch. In: Proceedings of the
  First Workshop on Bibliometric-enhanced Information Retrieval co-located with
  36th European Conference on Information Retrieval {(ECIR} 2014), Amsterdam,
  The Netherlands, April 13, 2014. {CEUR} Workshop Proceedings, vol.~1143, pp.
  37--44. CEUR-WS.org (2014), \url{http://ceur-ws.org/Vol-1143/paper5.pdf}

\bibitem{Fuhr:2018:CMI:3190580.3190586}
Fuhr, N.: Some common mistakes in ir evaluation, and how they can be avoided.
  SIGIR Forum  \textbf{51}(3),  32--41 (Feb 2018).
  \doi{10.1145/3190580.3190586}

\bibitem{DBLP:conf/jcdl/HienertKBZM19}
Hienert, D., Kern, D., Boland, K., Zapilko, B., Mutschke, P.: A digital library
  for research data and related information in the social sciences. In: 19th
  {ACM/IEEE} Joint Conference on Digital Libraries, {JCDL} 2019, Champaign, IL,
  USA, June 2-6, 2019. pp. 148--157. {IEEE} (2019).
  \doi{10.1109/JCDL.2019.00030}

\bibitem{ferro_continuous_2019}
Hopfgartner, F., Balog, K., Lommatzsch, A., Kelly, L., Kille, B., Schuth, A.,
  Larson, M.: Continuous {Evaluation} of {Large}-{Scale} {Information} {Access}
  {Systems}: {A} {Case} for {Living} {Labs}. In: Ferro, N., Peters, C. (eds.)
  Information {Retrieval} {Evaluation} in a {Changing} {World}, vol.~41, pp.
  511--543. Springer International Publishing, Cham (2019).
  \doi{10.1007/978-3-030-22948-1\_21}, series Title: The Information Retrieval
  Series

\bibitem{DBLP:journals/jdiq/JagermanBR18}
Jagerman, R., Balog, K., de~Rijke, M.: Opensearch: Lessons learned from an
  online evaluation campaign. J. Data and Information Quality  \textbf{10}(3),
  13:1--13:15 (2018). \doi{10.1145/3239575}

\bibitem{kramer_experimental_2014}
Kramer, A.D.I., Guillory, J.E., Hancock, J.T.: Experimental evidence of
  massive-scale emotional contagion through social networks. Proceedings of the
  National Academy of Sciences  \textbf{111}(24),  8788--8790 (Jun 2014).
  \doi{10.1073/pnas.1320040111}

\bibitem{lommatzsch2018newsreel}
Lommatzsch, A., Kille, B., Hopfgartner, F., Ramming, L.: Newsreel multimedia at
  mediaeval 2018: News recommendation with image and text content. In: Working
  Notes Proceedings of the MediaEval 2018 Workshop. CEUR-WS (2018)

\bibitem{radlinski_how_2008}
Radlinski, F., Kurup, M., Joachims, T.: How does clickthrough data reflect
  retrieval quality? In: Proceeding of the 17th {ACM} conference on
  {Information} and knowledge mining - {CIKM} '08. p.~43. ACM Press, Napa
  Valley, California, USA (2008). \doi{10.1145/1458082.1458092}

\bibitem{DBLP:conf/clef/SchaerSC20}
Schaer, P., Schaible, J., Castro, L.J.G.: Overview of lilas 2020 - living labs
  for academic search. In: Arampatzis, A., Kanoulas, E., Tsikrika, T.,
  Vrochidis, S., Joho, H., Lioma, C., Eickhoff, C., N{\'{e}}v{\'{e}}ol, A.,
  Cappellato, L., Ferro, N. (eds.) Experimental {IR} Meets Multilinguality,
  Multimodality, and Interaction - 11th International Conference of the {CLEF}
  Association, {CLEF} 2020, Thessaloniki, Greece, September 22-25, 2020,
  Proceedings. Lecture Notes in Computer Science, vol. 12260, pp. 364--371.
  Springer (2020). \doi{10.1007/978-3-030-58219-7\_24}

\bibitem{DBLP:conf/ecir/SchaerSM20}
Schaer, P., Schaible, J., M{\"{u}}ller, B.: Living labs for academic search at
  {CLEF} 2020. In: Jose, J.M., Yilmaz, E., Magalh{\~{a}}es, J., Castells, P.,
  Ferro, N., Silva, M.J., Martins, F. (eds.) Advances in Information Retrieval
  - 42nd European Conference on {IR} Research, {ECIR} 2020, Lisbon, Portugal,
  April 14-17, 2020, Proceedings, Part {II}. Lecture Notes in Computer Science,
  vol. 12036, pp. 580--586. Springer (2020).
  \doi{10.1007/978-3-030-45442-5\_75}

\bibitem{schuth_extended_2015}
Schuth, A., Balog, K., Kelly, L.: Extended overview of the living labs for
  information retrieval evaluation {(LL4IR)} {CLEF} lab 2015. In: Cappellato,
  L., Ferro, N., Jones, G.J.F., SanJuan, E. (eds.) Working Notes of {CLEF} 2015
  - Conference and Labs of the Evaluation forum, Toulouse, France, September
  8-11, 2015. {CEUR} Workshop Proceedings, vol.~1391. CEUR-WS.org (2015),
  \url{http://ceur-ws.org/Vol-1391/inv-pap8-CR.pdf}

\bibitem{de_solla_price_little_1963}
de~Solla~Price, D.J.: Little {Science}, {Big} {Science}. Columbia University
  Press, New York (1963)

\bibitem{DBLP:journals/corr/abs-2005-04474}
Voorhees, E.M., Alam, T., Bedrick, S., Demner{-}Fushman, D., Hersh, W.R., Lo,
  K., Roberts, K., Soboroff, I., Wang, L.L.: {TREC-COVID:} constructing a
  pandemic information retrieval test collection. CoRR  \textbf{abs/2005.04474}
  (2020), \url{https://arxiv.org/abs/2005.04474}

\bibitem{yang_critically_2019}
Yang, W., Lu, K., Yang, P., Lin, J.: Critically {Examining} the "{Neural}
  {Hype}": {Weak} {Baselines} and the {Additivity} of {Effectiveness} {Gains}
  from {Neural} {Ranking} {Models}. In: Proceedings of the 42nd {International}
  {ACM} {SIGIR} {Conference} on {Research} and {Development} in {Information}
  {Retrieval} - {SIGIR}'19. pp. 1129--1132. ACM Press, Paris, France (2019).
  \doi{10.1145/3331184.3331340}

\end{thebibliography}

\end{document}